# ACTIVITY INDUCED TURBULENCE IN DRIVEN ACTIVE MATTER

J. K. Bhattacharjee

School of Physical Sciences

Indian Association for the Cultivation of Sciences

Jadavpur, Kolkata 700032, India

and

T. R. Kirkpatrick

Institute of Physical Science and Technology

University of Maryland

College Park

Maryland 20742, U.S.A.

## Abstract

Turbulence in driven stratified active matter is considered. The relevant parameters characterizing the problem are the Reynolds number $\text{Re} = \bar{u}L/\nu$ and an active matter Richardson-like number, $R = |\zeta| S^2 L / D\bar{u}$. Here $\bar{u}$ is the mean velocity of flow, $L$ is the system size, $\nu$ is the kinematic viscosity, $\zeta$ is the activity coefficient, $S$ is the concentration gradient and $D$ is the active matter diffusion coefficient. In the *mixing* limit, $\text{Re} \gg 1, R \ll 1$, we show that the standard Kolmogorov energy spectrum law, $E(k) \propto k^{-5/3}$, is realized. On the other hand, in the *stratified* limit, $\text{Re} \gg 1, R \gg 1$, there is a new turbulence universality class with $E(k) \propto k^{-7/5}$. The crossover from one regime to the other is discussed in detail. Experimental predictions and probes are also discussed.

Over the last couple of decades a new variety of hydrodynamic instabilities and turbulence has been extensively studied [1-5] in the context of "active matter" hydrodynamics (e.g. bacteria swimming in a fluid). The earliest such studies involved "active nematics" [6-8] where a transition from a quiescent state to a spontaneously flowing state was predicted by Simha and Ramaswamy [9] and observed by Voituriez et al [10]. A large number of interesting results have also been obtained on turbulence in living fluids (e.g. Wensink et al [11]). The special feature of active matter hydrodynamics is that, because of its own energy source, active matter can introduce large additional stress terms in the Navier Stokes equation. The usual stress term for the velocity ($\vec{u}$) dynamics of an incompressible fluid is $T_{\alpha\beta} = -p\delta_{\alpha\beta} + \eta(u_{\alpha,\beta} + u_{\beta,\alpha})$ where '$p$' is the pressure, $u_{\alpha,\beta} = \partial_\beta u_\alpha$ and $\eta$ is the shear viscosity. The additional contribution to $T_{\alpha\beta}$ because of the active matter can take different forms depending on what is being studied. A particular form [12] of this extra term is reminiscent of the model 'H' among the different universality classes of dynamic critical phenomena [3-16]. The form of the contribution in the lowest non-trivial order allowed by symmetry considerations is a non-linear Burnett term [17-18] and can be written as

$$\Sigma_{\alpha\beta} = -\zeta\left(\partial_\alpha\phi\partial_\beta\phi - \frac{\delta_{\alpha\beta}}{3}(\nabla\phi)^2\right) \qquad (1)$$

where $\phi(\vec{r},t)$ is the concentration of the active matter and $\zeta$ is a constant which can be termed the activity coefficient. It should be noted that unlike inactive matter, the activity coefficient in dimensionless units need not be small. The dynamics of the system in the presence of statistical forcing for both velocity and concentration fields at large distance scales can be described by

$$\partial_t u_\alpha + u_\beta \partial_\beta u_\alpha = -\partial_\alpha p + \nu\nabla^2 u_\alpha + \partial_\beta \Sigma_{\alpha\beta} + f_\alpha \qquad (2a)$$

$$\partial_t \phi + u_\alpha \partial_\alpha \phi = D\nabla^2 \phi + g \qquad (2b)$$

The incompressibility condition is given by $\partial_\alpha u_\alpha = 0$. The statistical forces $\vec{f}(\vec{r},t)$ and $g(\vec{r},t)$ are Gaussian noise terms specified by zero mean and non-zero two point correlation function as in the pioneering work of De Dominicis and Martin [19] and followed up extensively by Yakhot and Orszag [20-21] and by Smith and Reynolds [22]. We focus on this particular model as it allows for the existence of an inertial range of wave numbers which do not contribute to the total energy input or dissipation as opposed to the active nematics [23] or systems with more complicated couplings in the fluid flow dynamics [24,25].

A driven version of the above model was introduced in Ref [26] where a fixed concentration gradient of the active matter was maintained across two parallel surfaces separated by a distance $L$. It was shown that for negative values of $\zeta$, an instability sets in when the dimensionless quantity $N = |\zeta| S^2 L^2 / D\nu$, which we will call the active matter Nusselt number, exceeds a critical value $N_c = 4\pi^2$. This instability is analogous to the convective instability in a fluid heated from below. It was shown by Das et al [27] that if the active matter Nusselt number is increased beyond $N_c$, then a cascade of period doubling bifurcations occur culminating in a chaotic state for $N \approx 50 N_c$.

In this work we show that a new universality class for turbulence can be induced in this system by tuning the active matter Nusselt number to values such that the parameter $R = N/\mathrm{Re} = |\zeta| S^2 L / D\bar{u}$ becomes much larger than unity. The quantity $\mathrm{Re}$ is the usual Reynolds number defined by $\mathrm{Re} = \bar{u}L/\nu$ where $\bar{u}$ is mean flow velocity and $\nu$ is the kinematic viscosity. Conventional Kolmogorov turbulence occurs for $\mathrm{Re} \gg 1, R \ll 1$. In this work we concentrate on the new regime characterized by $R \gg 1$ i.e. $N \gg \mathrm{Re}$ and find that

i) For R>>1, the Kolmogorov 5/3 law changes to a 7/5 law for the energy spectrum, i.e. $E(k) \propto k^{-7/5}$ in the inertial range of wave-numbers, setting up a new universality class of turbulence. This result is obtained in two ways : first via a scaling argument and then from the dynamical equations themselves.

ii) For a given value of R, the new spectrum is seen for wave-numbers $k \ll k_1$ and the Kolmogorov spectrum for $k \gg k_2$. The wave number $k_1$ is proportional to $\left(\zeta S^2 / \sigma\right)^{5/4}$ and the wave-number $k_2$ is proportional to $\left(\zeta S^2 / \sigma\right)^{3/2}$. Clearly for very large values of $\zeta S^2 / \sigma$ the spectrum is almost entirely the 7/5 variety, and for $\zeta S^2 / \sigma \ll 1$, the spectrum is almost entirely Kolmogorov like. This result is obtained from the dynamical equations.

It is interesting to note a special feature of this driven active model-H by contrasting it with the convective fluid system where a fluid layer is subjected to an adverse temperature gradient. For the latter case. using a three mode model ( a three dimensional dynamical system for the convecting fluid layer), Lorenz [28] found that as the gradient reaches a critical value the dynamics becomes extremely sensitive to initial conditions ( chaotic dynamics ). In real experiments however the dynamics evolves from stationary- in- time states to time periodic states ( Hopf bifurcation ) followed by more complicated time dependences before making a transition to a chaotic state. The subsequent discovery of dynamical systems showing period-doubling [29], intermittency [30], inherent instability [31] of more than two incommensurate frequency states had a very strong impact on hydrodynamic turbulence. These chaotic systems exhibited a complicated time dependence with the Fourier spectrum showing a continuous distribution of frequencies [32] but the spatial dynamics was ordered and characterized by only a few length scales and therefore not turbulent which requires a very large or infinite number of length scales.

Turbulence had long been a difficult problem to handle as it involved not only the complicated time dependence characteristic of chaotic systems but also involved an infinite number of length scales from the smallest (scale of viscous dissipation) to the largest (scale over which energy was supplied to the system e.g. the length $L$ introduced above). It has always been a challenge to find a physical system which, by changing a few parameters, can be taken from a zero velocity NESS state to a non-trivial steady state followed by a passage to chaos through a sequence of instabilities modelled by an appropriate low-dimensional dynamical system and then to a fully turbulent state by a further manipulation of the parameters.

To the best of our knowledge the driven active model H is the first example where one can study the passage to chaos ( increase the value of $N$ at a low Reynolds number ) and then study the passage to turbulence by either increasing the Reynolds number way beyond the Nusselt number or by increasing the Nusselt number way beyond the Reynolds number.

A turbulent state [33-35] in a homogeneous isotropic fluid is generally observed at very high values of the Reynolds number $\mathrm{Re}$. In steady state turbulence, the simplest and one of the most well- known results is the 5/3 law of Kolmogorov [36-37]. The steady state means that the amount of energy $(\varepsilon)$ introduced in unit time at large length scales is dissipated in unit time at the short viscous scales. In the intermediate length scales (smaller than the typical system size and larger than the viscous boundary layer thickness) the inducted energy cascades from the large length scales to small length scales at a constant rate $\varepsilon$ independent of the scale. This defines the inertial range. The total kinetic energy per unit mass of the system defines the energy spectrum $E(k)$ by the relation

$$E = \frac{1}{2V}\int d^3r \langle u_\alpha(\vec{r})u_\alpha(\vec{r})\rangle = \int \frac{d^3k}{2(2\pi)^3}\langle u_\alpha(\vec{k})u_\alpha(-\vec{k})\rangle = \int E(k)dk \qquad (3)$$

The assumption made by Kolmogorov was that the energy spectrum $E(k)$ is determined by $\varepsilon$ and $k$. A dimensional analysis leads to $E(k) \propto k^{-5/3}$ - the 5/3 law [36]. For the case of the stratified fluid when anisotropy is still not too big, the 5/3 changes to 11/5 as first predicted by Bolgiano [38] and Obukhov [39].

For the active stratified fluid, we begin by rewriting Eqs (2a) and (2c) in terms of variables which are centred around the non-equilibrium steady state ( NESS ) characterized by $\vec{u}=0$, constant pressure and a concentration distribution $\phi_0(\vec{r})$ which is written as $\phi_0(\vec{r}) = \phi_{00} + Sz$, where $\phi_{00}$ is a constant. We use the variable $\psi(\vec{r},t) = \phi(\vec{r},t) - \phi_0(\vec{r})$ and introduce the curl-free vector field $\vec{B} = \vec{\nabla}\psi$. Our interest being in the inertial range where the distance scale is always much larger than the viscous scale, the $\vec{B}$ field can be considered small ( small means small compared to S and hence in what follows the $B$ field that will be written is actually $B/S$ and we can linearize in it to arrive at the system ( we write the fields in wave-number space to facilitate calculations later)

$$\partial_t u_\alpha(k) + M_{\alpha\beta\gamma}(k)\int \frac{d^3p}{(2\pi)^3} u_\beta(p)u_\gamma(k-p) = -iS^2\zeta\left[\delta_{\alpha 3}k_\beta B_\beta(k) - k_3 B_\alpha(k)\right] - \eta k^2 u_\alpha(k) + f_\alpha(k)$$
(4a)

$$\partial_t B_\alpha + ik_\alpha \int \frac{d^3p}{(2\pi)^3} u_\beta(p)B_\beta(k-p) = -Dk^2 B_\alpha(k) + iS^2 u_3 k_\alpha + ik_\alpha g(k) \qquad (4b)$$

The coefficient $M_{\alpha\beta\gamma}(k)$ of the nonlinear term in the velocity dynamics is $i\left[k_\beta P_{\alpha\gamma}(k) + k_\gamma P_{\alpha\beta}(k)\right]/2$ and the incompressibility condition is $k_\alpha u_\alpha(k)=0$. The conserved quantity in the inviscid, unforced limit ( $\eta = D = \vec{f} = g = 0$ ) is seen to be the total energy per unit mass of the active fluid

$$E = \int \frac{d^3p}{2(2\pi)^3}\left[u_\alpha(p)u_\alpha(-p) + \zeta S^2 B_\alpha(p)B_\alpha(-p)\right] \qquad (5)$$

The difference from the Kolmogorov situation of Eq (3) is that the energy, in addition to the usual kinetic energy term, has a term which we label as the potential energy. It is the total energy given above which is dissipated at large values of the wave-vector by the viscosity $\eta$ and the concentration diffusion $D$. To keep the total energy constant in the presence of the dissipative terms we add the small (in wave-vector) scale noise terms $\vec{f}(k,t)$ and $g(k,t)$. The rate of energy flow across a given wave-number in the inertial range is given by

$$\varepsilon(k) = \int_0^k \frac{d^3 p}{(2\pi)^3}\left[\dot{u}_\alpha(p)u_\alpha(-p) + \zeta S^2 \dot{B}_\alpha(p)B_\alpha(-p)\right] + c.c. \qquad (6)$$

In the above $c.c.$ stands for the complex conjugate. When $\zeta S^2 \ll 1$, the second term is negligible and the energy spectrum $E(k)$ is the Kolmogorov variety. Our primary interest is in the limit $\zeta S^2 \gg 1$. In this case we have the interesting situation where the energy spectrum, which is by definition the kinetic energy, is dominated by the flux of the potential energy. Hence the Kolmogorov scaling argument must be redone.

To do this we need to find the scaling dimension of the $B$-field. It is simplest to do so by considering Eq. (4a) in real space and looking at the linear terms where the acceleration is driven by the gradient of the $B$-field. The scaling dimension is the dimension that leaves the real space version unchanged when the length scale changes by a factor $\alpha$ i.e. when we consider the transformation $l \to \alpha l$. If time scales under this transformation as $\alpha^z$, then the acceleration scales as $\alpha^{1-2z}$ and hence $B$ as $\alpha^{2-2z}$. The quantity $\varepsilon(k)$ consequently scales $\alpha^{4-5z}$. For the flux to be $k$-independent we need $z = 4/5$. The energy spectrum $E(k)$ has the dimension $L^3/T^2$ and thus scales as $l^{3-2z}$ which is $k^{2z-3}$. Using $z = 4/5$, we have the spectrum $E(k) \propto k^{-7/5}$ for $\zeta \gg 1$. Denoting the $B$-field flux by $\varepsilon_B$, we have the kinetic energy spectrum given by

$$E(k) = K'\varepsilon_B^{2/5} k^{-7/5} \qquad (7)$$

In Eq (7) above the universal numerical constant K' is the analogue of the Kolmogorov constant for the usual turbulence. Note that if $S = 0$, the same sort of arguments lead to the Kolmogorov result $E(k) \propto k^{-5/3}$.

Next we show that the dynamics specified by Eqs (4a)-(4b) is consistent with the conclusion above. This can be seen by working directly in the approximation $\zeta S^2 \gg 1$, where the nonlinearity in Eq (4a) is overwhelmed by the $\zeta$-containing term and thus

$$u_\alpha(k,t) = -i\zeta S^2 (\delta_{\alpha 3} k_\beta - \delta_{\alpha\beta} k_3)\int_0^t dt'\, G_u(k, t-t') B_\beta(k,t'),$$ where $G_u(k,t)$ is the dressed

propagator for the velocity field. To obtain the dynamics of $B_\alpha(k,t)$, we write the solution of Eq.(4b) as

$$B_\alpha(k,t) = i\int_0^t dt' G_B(k,t-t') \int \frac{d^3p}{(2\pi)^3} k_\alpha u_\beta(p,t') B_\beta(k-p,t') \qquad (8)$$

The dressed propagator $G_B(k,t)$ is written in wave-number frequency space by the usual Dyson equation $G_B^{-1}(k,\omega) = G_0^{-1}(k,\omega) + \Sigma_B(k,\omega)$ with the dressed one-loop self–energy given by

$$\Sigma_B(k,\omega) = k^2 \int\int \frac{d\omega'}{2\pi} \frac{d^3p}{(2\pi)^3} G_B(p,\omega') C_u(k-p,\omega-\omega') \qquad (9)$$

In the above equation, $C_u(k,\omega)$ is the velocity correlation function. We now match the scaling dimensions of either side. The inverse of the Greens function and the self-energy scale as the frequency and hence behave as $k^z$. The equal time $B$–field correlation function $\int d\omega C_B(k,\omega)$ is taken to scale as $k^{-n}$. The relation between the velocity and the $B$–field shown above gives the scaling dimension of $\int d\omega C_u(k,\omega)$ as $k^{-n+2-2z}$. The matching of scaling properties of the two sides of Eq. (9) yields $4z = 7-n$.

We need the energy transfer rate $\varepsilon(k)$ of Eq. (6) at the lowest dressed order of perturbation theory in the $\zeta \gg 1$ limit. We drop the first term on the right hand side of Eq (6) and use the nonlinear term of Eq (4c) to obtain

$$\varepsilon(k) = -2i\zeta \langle \int_0^k \frac{d^3p}{(2\pi)^3} p_\alpha \int \frac{d^3q}{(2\pi)^3} B_\alpha(-p,t) B_\beta(p-q,t) u_\beta(q,t) \rangle \qquad (10)$$

where the angular bracket stands for the average over the random forcing term in Eq (4b). Because of the dynamics involved in the term $B_\beta(p-q,t)$ of Eq (10) above, we write from Eq (4b)

$$B_\beta(p-q,t) = i\int_0^t dt' G_B(p-q,t-t') \int \frac{d^3l}{(2\pi)^3} (p-q)_\beta u_\gamma(l,t') B_\gamma(p-q-l,t') \qquad (11)$$

We substitute for $B_\beta(p-q,t)$ in Eq (10) from Eq.(11) and use the connection between $u_\alpha$ and $B_\beta$ given above Eq (8) to arrive at

$$\varepsilon(k) = \int_0^k \frac{d^3p}{(2\pi)^3} \int_0^t dt' \int \frac{d^3q}{(2\pi)^3} \int \frac{d^3l}{(2\pi)^3} p_\alpha (p-q)_\beta G_B(p-q,t-t') \langle B_\alpha(-p,t) B_\gamma(p-q-l,t') \rangle \langle u_\beta(q,t) u_\gamma(l,t') \rangle$$
(12)

Using the relation between the velocity field and $B$–field, and the Kolmogorov condition of $\varepsilon(k)$ being independent of $k$ leads to $3z + 2n = 10$. Combining with the other relation between the two exponents found from the self energy consistency leads to $z = 4/5, n = 19/5$. The relation between the velocity and the B-field now gives $E(k)$ with $E(k) \propto k^{-7/5}$ once again. In terms of the wave-number and energy flux, we have the new scaling regime where the energy spectrum follows a 7/5 law.

Having established that for $S^2 \zeta \gg 1$, the turbulent kinetic energy spectrum follows a new scaling law with the dynamical equations, we now ask the question of how the crossover from Kolmogorov variety to this new variety of turbulence occurs as the control parameter $N$ is varied. We begin by recalling the Heisenberg [40,41] theory of turbulence and the extension of it by Chandrasekhar [42] to account for the crossover from the Kolmogorov regime to the viscosity dominated regime. Heisenberg began by noting that the energy transfer due to viscosity from small to large wave-numbers across a given wave-number $k$ in Eq (4a) is given by $-\eta \int_0^k \frac{d^3p}{(2\pi)^3} \langle u_\alpha(p) u_\alpha(-p) \rangle = -\eta \int_0^k dp\, p^2 E(p)$. In analogy with this he decided to write the transfer caused by the non-linear term in Eq (4a) as a non-local effective viscosity term given by $-\eta_{eff}(k) \int_0^k dp\, p^2 E(p)$ Since the transfer occurs to all scales larger than $k$, Heisenberg used dimensional arguments to write the $\eta_{eff}$ (eddy viscosity) in terms of $E(p)$ and $p$ as $\eta_{eff}(k) = \int_k^\infty \frac{dp}{p} \sqrt{\frac{E(p)}{p}}$.

In our case in addition to the viscous dissipation there is an additional term from the concentration diffusion. The total dissipation for us is $\int \frac{d^3p}{(2\pi)^3} \left[ \eta \langle u_\alpha(p) u_\alpha(-p) \rangle + D\zeta S^2 \langle B_\alpha(p) B_\alpha(-p) \rangle \right]$. The first term of the dissipation is automatically $-\eta \int E(p) dp$ and is treated as above. We write the second term in the dissipation in terms of a $D_{eff}$ analogous to the $\eta_{eff}$ above and express the $B$-field correlation function in terms of $E(p)$ and $p$ by a dimensional analysis.

This makes its contribution of the form $v_{eff} \frac{\zeta S^2}{\sigma} \int p^3 E^2(p) dp$ where $\sigma = v_{eff} / D_{eff}$ is the turbulent Prandtl number and is assumed to be a simple number. Absorbing all numerical factors of $O(1)$ in $v_{eff}$ and $\sigma$, we write the energy flux in the inertial range as

$$\varepsilon(k) = \eta_{eff}(k) \left[ \int_0^k E(p) p^2 dp + \frac{\zeta S^2}{\sigma} \int_0^k E^2(p) p^3 dp \right] = \eta_{eff}(k) y(k) \quad (13)$$

In terms of a function $g(k)$ is defined by the relation $g(k) = k \frac{dy}{dk} = E(k)k^2 + \zeta S^2 E^2(k) k^3 / \sigma$, we get

$$\frac{2\zeta}{\sigma} k E(k) = \sqrt{1 + 4 \frac{g(k)}{k} \frac{\zeta S^2}{\sigma}} - 1 \quad (14)$$

Substituting for $\eta_{eff}(k)$ in Eq (14) from Eq (13) and imposing the scale independent energy flux condition i.e. $\varepsilon(k)$ is independent of $k$ gives

$$\int_k^\infty \sqrt{\frac{E(p)}{p^3}} dp = \frac{y(k)}{k^2 g(k)} \left( \sqrt{1 + \frac{4\zeta S^2}{\sigma} \frac{g(k)}{k^2}} - 1 \right)^{1/2} \quad (15)$$

The most relevant information about the crossover can be extracted from Eq.(15) itself. The vital point about the crossover is that it is not determined by $\zeta S^2 / \sigma$ alone, but by the quantity $\xi = \zeta S^2 g(k) / \sigma k^2$, showing that along with $\zeta / \sigma$, the wave-number plays a very important role in the crossover. For $\xi \ll 1$, the energy spectrum is Kolmogorov, while for $\xi \gg 1$ it is the new variety $E(k) \propto k^{-7/5}$ established in Eq.( 7 ).

It is easier to use the variable $y(p)$ instead of $p$ in Eq. (15). Doing this on the l.h.s and using the limit $Ri \to 0 (\xi \ll 1)$ on the right, we get after taking a derivative with respect to $y(k)$, the flow equation

$$\frac{dg}{dy} - 4 \frac{g}{y} + 4 = 0 \quad (16)$$

The solution is the Kolmogorov spectrum in the inertial range [ 42]. Since the above equation corresponds to a solution $g(k) \propto k^{4/3}$, the condition $\xi \ll 1$ holds for wave-numbers $k$ which are larger than $k_2 \propto \left( \zeta S^2 / \sigma \right)^{3/2}$.

The limit $\xi \gg 1$, on the other hand, leads to

$$\frac{dg}{dy} - \frac{8g}{3} = -2 \qquad (17)$$

Since $\frac{dy}{dk} = \frac{g}{k}$ implies $\ln k = \int \frac{dy}{g}$, we can integrate the above equation to obtain $g(k) = \frac{6k^{6/5}}{5(1+\beta k^2)^{8/5}}$ where $\beta$ is a constant of integration. Our concern being with the inertial range, the wave-number $k$ can be considered smaller than the scale $\beta^{-1/2}$ ( dissipative scale for the concentration fluctuations ) and we have $g(k) \propto k^{6/5}$. From Eq.(15), this yields $E(k) \propto k^{-7/5}$ for $\zeta S^2/\sigma \gg 1$. More accurately, we need $\frac{\zeta S^2}{\sigma} \frac{g(k)}{k^2} \gg 1$ which requires $k \ll k_1$ where $k_1 \propto (\zeta S^2/\sigma)^{5/4}$. Clearly for $\zeta S^2/\sigma > 1$, we have $k_2 > k_1$. Hence, the final picture that emerges is that for wave-numbers $k < k_1$, the spectrum is purely of the $k^{-7/5}$ variety and for $k > k_2$ it is of the Kolmogorov variety. The region between $k_1$ and $k_2$ corresponds to the crossover from one scaling to another.

We conclude by pointing out that if measurements are done in real space then the relevant quantity is the two point correlation function

$$S_2(r) = \langle (u(x+r)-u(x))^2 \rangle = 4\int_0^\infty E(k)\left[1 - \frac{\sin kr}{kr}\right]dk \quad.$$

The scaling behaviour of the energy spectrum leads to the conclusion that at short length scales the correlation function $S_2(r)$ will behave as $r^{2/3}$ ( Kolmogorov), while at larger length scales the scaling relation will be $S_2(r) \propto r^{2/5}$. In the case of buoyancy driven turbulence the Kolmogorov regime is obtained at large length scales and the short scales lead to the Bolgiano-Obukhov scaling of $r^{6/5}$. This crossover was found numerically and experimentally by Kunnen etal [ 43 ]. For this case of driven active matter turbulence we predict that the Kolmogorov behaviour will be seen at short spatial scales and a $r^{2/5}$ behaviour at large length scales.

We end by pointing out that the recent experimental advances in the technique for generating linear concentration gradient of chemo-attractants in a channel [44-46] makes it possible to test our predictions. It is possible that the early experiments would find it easier to probe the correlations of the B-field rather than the velocity field. To this end, we point out that a simple extension of the scaling arguments leads to the "potential energy" spectrum $E_B(k)$ defined by $\int E_B(k)dk = V^{-1}\int d^3r \langle B_\alpha B_\alpha \rangle$ shows that $E_B(k) \propto k^{-7/5}$ if the kinetic energy flux dominates ( Kolmogorov) and $E_B(k) \propto k^{-9/5}$ if the bacterial gradient flux dominates. The crossover is now reversed. It is Kolmogorov mechanism at the higher wave- numbers. In real space the two point function goes as $r^{2/5}$ at shorter scales crossing over to $r^{4/5}$ at larger scales.